\documentclass[12pt,a4paper]{article}

\usepackage{graphicx,cite}

\newcommand{\figwidth}{12cm}
\newcommand{\figheight}{7.0cm}

\newcommand{\nn}{\nonumber}

\newcommand{\IM}{\mbox{\rm Im}}

\newcommand{\mev}{\mbox{\rm MeV}}
\newcommand{\gev}{\mbox{\rm GeV}}
\newcommand{\eqn}[1]{(\ref{#1})}

\newcommand{\ep}{\epsilon}

\newcommand{\pslh}{p\!\!\!\!\!\not\,\,\,}

\newcommand\lsim{\mathrel{\rlap{\lower4pt\hbox{\hskip1pt$\sim$}}
    \raise1pt\hbox{$<$}}}
\newcommand\gsim{\mathrel{\rlap{\lower4pt\hbox{\hskip1pt$\sim$}}
    \raise1pt\hbox{$>$}}}
\newcommand{\quarkkond}{\langle \bar{q}q \rangle}
\newcommand{\strangekond}{\langle \bar{s}s \rangle}
\newcommand{\gluonkond}{\langle \frac{\alpha_s}{\pi}FF \rangle}
\newcommand{\mixedkond}{\langle g_s\bar{q}\sigma Fq \rangle}
\newcommand{\strangemixedkond}{\langle g_s\bar{s}\sigma F s \rangle}

\begin{document}

\begin{titlepage}

\begin{flushright}
{\small\sf IFIC/04-15\\FTUV/04-0415} \\[15mm]
\end{flushright}

\begin{center}
{\Large\bf Pentaquark and diquark-diquark clustering:\\
a QCD sum rule approach}\\[10mm]

{\normalsize\bf Markus Eidem\"uller}\\[4mm]

{\small\sl Departament de F\'{\i}sica Te\`orica, IFIC,
           Universitat de Val\`encia -- CSIC,}\\
{\small\sl Apt. Correus 22085, E-46071 Val\`encia, Spain} \\[17mm]
\end{center}

\begin{abstract}
\noindent
In this work we study the $\Theta^+(1540)$ in the framework of QCD sum rules
based on $(ud)^2\bar{s}$ diquark clustering as suggested by Jaffe and
Wilczek. Within errors, the mass of the pentaquark is compatible with the
experimentally measured value. The mass difference between the $\Theta^+$ and
the pentaquark with the quantum numbers of the nucleon amounts to
70 MeV, consistent with the interpretation of the $N(1440)$ as a pentaquark.
\end{abstract}

\vfill


\end{titlepage}




\noindent
Recently, several experiments
\cite{Nakano:2003qx,Barmin:2003vv,Stepanyan:2003qr,Barth:2003ja,
Kubarovsky:2003fi,Airapetian:2003ri,Asratyan:2003cb,
Aleev:2004sa,:2004kn,Abdel-Bary:2004ts}
have observed a new baryon resonance
$\Theta^+(1540)$ with positive strangeness. Therefore it 
requires an $\bar{s}$ and has a 
minimal quark content of five quarks. This discovery has triggered 
an intense experimental and theoretical activity to clarify the quantum 
numbers and to understand the structure of the pentaquark state. 
The $\Theta$ has the third component of isospin zero 
and the absence of isospin partners suggests strongly that the $\Theta$ 
is an isosinglet what we also assume in this work. 
A puzzling characteristics of the $\Theta$ is its narrow width below
15 MeV. A suggestive way to explain the small width is by the assumption of
diquark clustering. The formation of diquarks presents an important concept
and has direct phenomenological impact \cite{Diquarks}.
Two models have been proposed 
based on the strong attraction of the $(ud)$-diquarks:
one by Karliner and Lipkin \cite{Karliner:2003sy} 
where the pentaquark is described as
diquark-triquark system in a non-standard colour representation. The other one 
is due to Jaffe and Wilczek \cite{Jaffe:2003sg,Jaffe:2004zg} and 
describes the $\Theta$ as
bound state of an $\bar{s}$ with two highly correlated $(ud)$-diquarks.
In this work we investigate the second approach by Jaffe and Wilczek in 
the framework of QCD sum rules. 
In principle, as was discussed in \cite{Jennings:2003wz}, 
even a mixing between the two
states could be possible. However, an estimation of such a
potential mixing would require a detailed investigation of the model by
Karliner and Lipkin.
The basis of the sum rules was laid
in \cite{SRbasis} and their extension to baryons was developed in
\cite{SRbaryons}.
The assumptions of the model are
incorporated by an appropriate current. 
Since the sum rules are directly based on QCD and keep the analytic dependence
on the input parameters, they can help to differentiate between the
models and to test their features.
The relevance of the diquark picture within the context of the 
sum rules was shown in \cite{SRdiquarks}.
Several sum rule investigations for the pentaquark already exist 
\cite{Sugiyama:2003zk,Zhu:2003ba,Matheus:2003xr,Huang:2003bu}
which, however, are based on different models or currents.
The diquark models for the pentaquark have also been investigated 
within other approaches \cite{Pentaapproaches}.

In the model by Jaffe and Wilczek
the $(ud)$-diquarks have zero spin and are in a $\bar{3}_c$
and $\bar{3}_f$ representation of colour and flavour. In order to combine with
the antiquark into a colour singlet, the two diquarks must combine into a
colour 3. The diquark-diquark wavefunction is antisymmetric and has angular
momentum one. This combines with the spin of the $\bar{s}$ to total 
angular momentum 1/2 and results in positive parity. 
In \cite{Jaffe:2003sg} it was suggested to interpret the Roper resonance 
$N(1440)$ as $(ud)^2\bar{d}$ pentaquark state and
we will study this resonance at the end of our analysis.


The basic object in our sum rule analysis is the two-point correlation
function 
\begin{equation}
 \label{eq:PiDef}
\Pi(p) = i \int d^4 x\ e^{ipx} \langle 0|T\{\eta(x)\bar{\eta}(0)\}|0\rangle\,,
\end{equation}
where $\eta(x)$ represents the interpolating field of the 
pentaquark under investigation. 

The diquarks have a particularly strong attraction in the flavour antisymmetric
$J^P=0^+$ channel. Thus the current contains two diquarks of the form
\begin{equation}
 \label{eq:Diquark}
{\cal Q}^c(x) = \ep^{abc}\,Q_{ab}(x) = \ep^{abc}\,[u_a^T C\gamma_5 d_b](x)\,.
\end{equation}
$C$ denotes the charge conjugation matrix.
The two diquarks must be in a $p$-wave to satisfy Bose statistics.
Therefore the current contains a derivative to generate one unit of angular
momentum. The diquarks couple to a $3_c$ in colour to form the current
\begin{equation}
 \label{eq:Current}
\eta(x)= \left(\ep^{abd}\delta^{ce}-\ep^{abc}\delta^{de}\right)
\left[Q_{ab}\left(D^\mu Q_{cd}\right)-\left(D^\mu Q_{ab}\right)Q_{cd}
\right]\gamma_5 \gamma_\mu C \bar{s}_e^T\,,
\end{equation}
where the covariant derivative for the $\bar{3}_c$ is given by
$D^\mu=\partial^\mu-ig \lambda_l^\dagger A^{\mu\,l}$ \cite{Jaffe:2004zg}.
The parity is positive.
This current has a different structure than the current in 
\cite{Matheus:2003xr} which contains no derivative to produce the angular
momentum between the diquarks. Inserting the current and
neglecting higher orders in the strong coupling constant the correlator is
given by 
\begin{eqnarray}
 \label{eq:Correlator}
\Pi(x)&=& \langle 0|T\{\eta(x)\bar{\eta}(0)\}|0\rangle\ 
= \left[\gamma_5\gamma^\mu C S^{(s)\,T}_{e'e}(-x) C \gamma^\nu\gamma_5\right] 
T_{\mu\nu}^{ee'}(x)\,, \nn\\ 
T_{\mu\nu}^{ee'}(x) &=& \left(\ep^{abd}\delta^{ce}-\ep^{abc}\delta^{de}\right)
\left(\ep^{a'b'd'}\delta^{c'e'}-\ep^{a'b'c'}\delta^{d'e'}\right)\nn\\
&\times&\left[-\partial^{(cd)}_\mu \partial^{(c'd')}_\nu 
+\partial^{(cd)}_\mu \partial^{(a'b')}_\nu
+\partial^{(ab)}_\mu \partial^{(c'd')}_\nu 
-\partial^{(ab)}_\mu \partial^{(a'b')}_\nu\right]\nn\\
&\times&\Big\{
\langle\gamma_5 S_{bb'}(x) \gamma_5 C S_{aa'}^T (x)C\rangle
\langle\gamma_5 S_{dd'}(x) \gamma_5 C S_{cc'}^T (x)C\rangle\nn\\
&+&\langle\gamma_5 S_{bd'}(x) \gamma_5 C S_{ac'}^T (x)C\rangle
\langle\gamma_5 S_{db'}(x) \gamma_5 C S_{ca'}^T (x)C\rangle\nn\\
&-&\langle\gamma_5 S_{bd'}(x) \gamma_5 C S_{cc'}^T (x)C
\gamma_5 S_{db'}(x) \gamma_5 C S_{aa'}^T (x)C\rangle\nn\\
&-&\langle\gamma_5 S_{bb'}(x) \gamma_5 C S_{ca'}^T (x)C
\gamma_5 S_{dd'}(x) \gamma_5 C S_{ac'}^T (x)C\rangle
\Big\}\,,
\end{eqnarray}
with $\partial^{(ab)}_\mu=\partial^{(a)}/\partial
x^\mu+\partial^{(b)}/\partial x^\mu$ and the upper colour index indicates the
propagator on which the derivative is acting. $S(x)$ and $S^{(s)}(x)$ are the
light and strange quark propagators, respectively.
The quark propagator has been evaluated in the presence of quark and gluon
condensates in \cite{Novikov:1985gd,Yang:1993bp,Matheus:2003xr}, 
where the explicit expressions can be found.
Using the following Lorentz decomposition for 
$T_{\mu\nu}^{ee'}=\delta^{ee'} T_{\mu\nu}/3$,
\begin{equation}
 \label{eq:Lorentz}
T_{\mu\nu} = g_{\mu\nu}f_1(x^2)+ x_\mu x_\nu f_2(x^2) \,,
\end{equation}
the functions $f_1(x^2)$ and $f_2(x^2)$ are determined to
\begin{eqnarray}
 \label{eq:fCoeff}
f_1(x^2) &=& \frac{576}{\pi^8 x^{14}} - \frac{240 m_q^2}{\pi^8 x^{12}}
+ \frac{24 m_q^4-64 \pi^2 m_q \quarkkond + \frac{29 \pi^2}{8} \gluonkond}{\pi^8
  x^{10}}\nn\\
&&+ \frac{12 m_q^3 \quarkkond -4 m_q \mixedkond -16 \pi^2 \quarkkond^2}{\pi^6 x^{8}}
+O(1/x^6)\,,  \nn\\
f_2(x^2) &=& -\frac{1152}{\pi^8 x^{16}} + \frac{576 m_q^2}{\pi^8 x^{14}}
+ \frac{-48 m_q^4+256 \pi^2 m_q \quarkkond - \frac{61\pi^2}{4} \gluonkond}
{\pi^8 x^{12}}\nn\\
&&+ \frac{-32 m_q^3 \quarkkond + 32 m_q \mixedkond 
+128 \pi^2 \quarkkond^2}{\pi^6 x^{10}} +O(1/x^8)\,.  
\end{eqnarray}
The colour non-diagonal part of $T_{\mu\nu}^{ee'}$ vanishes for the considered
orders. 
In momentum space the correlator can be parametrised as
\begin{equation}
 \label{eq:MomCorr}
\Pi(p)=\pslh \Pi^{(p)}(p^2) + \Pi^{(1)}(p^2)\,.
\end{equation}
To obtain the phenomenological side we insert intermediate baryon states with
the corresponding quantum numbers. The matrix element of the $\Theta$ is
parametrised by 
\begin{equation}
 \label{eq:Matrix}
\langle 0|\eta(0)|\Theta(p)\rangle = f_\Theta\cdot u(p)\,.
\end{equation}
Since no experimental information on higher pentaquark states is available we
make the assumption of quark-hadron duality and approximate the higher states
by the perturbative spectral density above a threshold $s_0$. In fact, the
uncertainty on $s_0$ will be one of the dominant errors in the sum rule
analysis.

In order to suppress the higher dimensional condensates and to reduce the
influence of the higher resonances we employ a Borel
transformation defined by
\begin{equation}
 \label{eq:Borel}
\hat{{\cal B}}_M= \lim_{Q^2,n\to\infty} \frac{(-Q^2)^n}{\Gamma(n)}
\left(\frac{d}{dQ^2}\right)^n\,, \quad M^2=\frac{Q^2}{n}\ \mbox{fixed}\,,
\end{equation}
with $Q^2=-p^2$. 
As in \cite{Zhu:2003ba,Matheus:2003xr} we now concentrate on the chirality even part
$\Pi^{(p)}$ in eq. \eqn{eq:MomCorr} which contains the leading order term from
the operator product expansion.
The spectral density $\rho(s)=1/\pi \IM\Pi^{(p)}(s+i\ep)$ has the form
\begin{equation}
 \label{eq:}
\rho(s)= a_6 s^6 + a_5 s^5 + a_4 s^4+ a_3 s^3+\ldots
\end{equation}
The coefficients $a_i$ can easily be obtained from the results of 
eqs. \eqn{eq:Correlator} and \eqn{eq:fCoeff} by inserting the strange quark
propagator and performing a Fourier transformation.
The theoretical moments are then given by
\begin{eqnarray}
 \label{eq:TheoMoments}
\hat{\Pi}(M^2) &=& \hat{\cal B}_M\ \Pi(Q^2) 
= \int_0^\infty ds \ \frac{\rho(s)}{M^2} e^{-s/M^2}=  a_6 \Gamma(7)
(M^2)^6\nn\\ 
&& + a_5 \Gamma(6)  (M^2)^5 + a_4 \Gamma(5) (M^2)^4+ a_3 \Gamma(4) (M^2)^3 
+\ldots 
\end{eqnarray}
Transferring the continuum contribution to the theoretical side and taking a
logarithmic derivative with respect to $-1/M^2$, one obtains the sum rule for
the mass of the pentaquark,
\begin{eqnarray}
 \label{eq:Mass}
m_\Theta^2 &=& \frac{\sum\limits_{k=0}^{k=3} a_{6-k} \Gamma(8-k)
(M^2)^{8-k} E_{7-k}}
{\sum\limits_{k=0}^{k=3} a_{6-k} \Gamma(7-k) (M^2)^{7-k} E_{6-k}}\,,
\end{eqnarray}
where $E_\alpha=1-\Gamma(\alpha+1,s_0/M^2)/\Gamma(\alpha+1)$.


A basic input for the sum rule analysis is the Borel parameter $M$.
The sum rule should be stable with respect to $M$ to allow a reliable
determination of the pentaquark mass. 
For large values of $M$ the operator product expansion 
converges well, however, for small $M$ the expansion becomes problematic and
thus we restrict the range of the Borel parameter to $M\gsim 1.6\ \gev$.
Small $M$ suppress the phenomenological continuum part which becomes 
very dominant for large $M$. Therefore we employ a sum rule window of
$2.5\ \gev^2 < M^2 < 4.0\ \gev^2$.

As input parameters in our analysis we use 
$m_s= 0.15\ \gev$, $\quarkkond=-(0.267\pm 0.018\ \gev)^3$, 
$\strangekond = (0.8 \pm 0.2) \quarkkond$, 
$\strangemixedkond=M_0^2 \strangekond$ with $M_0^2=(0.8\pm 0.2)\ \gev^2$,
and $\gluonkond= 0.024 \pm 0.012\ \gev^4$ \cite{inputparameters}.
For the continuum threshold we use a central value of 
$s_{0}=(1.54+0.35\ \gev)^2$. Thus the continuum starts 350 MeV above the 
measured pentaquark mass. This difference should roughly correspond to one
radial excitation \cite{Zhu:2003ba} and represents 
a typical value for sum rule analyses with light
quarks as degrees of freedom \cite{SRbasis}.
\begin{figure}
\begin{center}
\includegraphics[height=\figheight,width=\figwidth,angle=0]{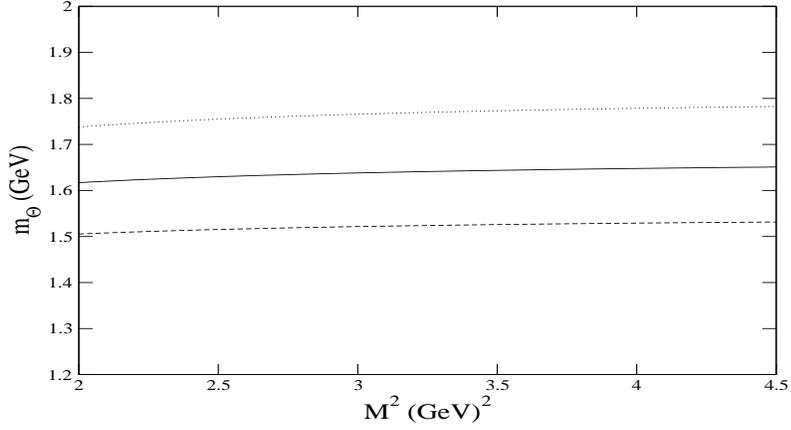}
\caption{\label{fig:1} $m_\Theta$ as a function of the Borel parameter $M^2$
for different $s_0=3.5\ \gev^2$ (solid), $s_0=4.1\ \gev^2$ (dotted) and 
$s_0=3.0\ \gev^2$ (dashed).}
\end{center}
\end{figure}
Fig. \ref{fig:1} shows the mass as a function of the Borel parameter
$M^2$. The sum rule has a good stability with respect to $M$.
As central value for the pentaquark mass we obtain $m_\Theta=1.64\ \gev$.
The two most important sources of the error are the choice of the continuum
threshold and the convergence of the operator product expansion. Since
we have substituted the phenomenological spectral density, using the assumption
of quark hadron duality, by the perturbative expansion,
the uncertainty on $s_0$
reflects the missing knowledge of the experimental cross section for higher
energies. 
To estimate the error on $m_\Theta$ we vary $s_0$ between
$3.0\ \gev^2 < s_0 < 4.1\ \gev^2$. 
In fig. \ref{fig:1} we have also plotted the change of $m_\Theta$ with the
continuum threshold from which we obtain an error of 
$\Delta m_\Theta \approx 125\ \mev$.
More phenomenological information would be essential to reduce 
this kind of error.
\begin{figure}
\begin{center}
\includegraphics[height=\figheight,width=\figwidth,angle=0]{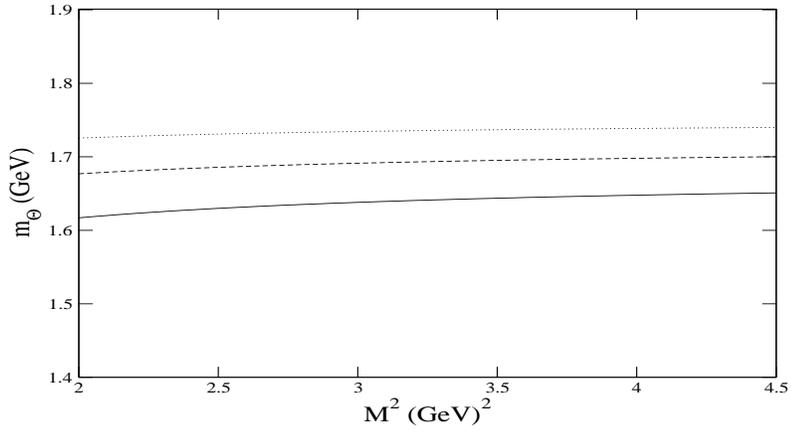}
\caption{\label{fig:2} $m_\Theta$
for different orders of the OPE, using only the leading order perturbative
expansion (dotted), with dimension 4 condensates (dashed) and including the
condensates of dimension 6 (solid).} 
\end{center}
\end{figure}
To estimate the dependence of the sum rules on the OPE we successively remove
the different orders. Fig. \ref{fig:2} shows the
convergence of the pentaquark mass including the condensate contributions
up to a specific power.
The inclusion of the higher condensates lowers the mass.
Using only the leading order perturbative result the central value is about 100
MeV larger than the full result. 
We have not included an extra graph for the term $\propto a_5$
since this contribution is proportional to the light quark masses and their
influence on the analysis can be neglected. The four-dimensional condensates
lower the leading order result by about 50 MeV and the condensates of
dimension 6 by another 50 MeV. We assume that a reasonable error estimate
from the OPE would be $\Delta m_\Theta \approx 75\ \mev$. 
Furthermore, contributions to the error also arise from the other input 
parameters which we vary in the ranges presented above. 
As it turns out, their influence on the value of $m_\Theta$ is
small compared to the errors from the continuum threshold 
and the convergence of the OPE.
Adding the errors quadratically our final result reads
\begin{equation}
 \label{eq:PentaMass}
m_\Theta = 1.64 \pm 0.15 \ \gev.
\end{equation}
In \cite{Jaffe:2003sg} Jaffe and Wilczek suggested to interpret the 
Roper resonance as $(ud)^2\bar{d}$ pentaquark state. 
One can then perform a similar analysis for the $N(1440)$ as has been done
for the $\Theta$ by substituting the $\bar{s}$ antiquark by a 
$\bar{d}$ antiquark. 
As central value for the continuum threshold we choose, as in the $\Theta^+$
case, a value of 350 MeV above the ground state mass. For the error range we
use $2.7\ \gev^2 < s_{0N} < 3.8 \ \gev^2$.
Performing a sum rule analysis for the $N$ with the above given parameters,
we obtain a mass of $m_N=1.57 \pm 0.15\ \gev$.
\begin{figure}
\begin{center}
\includegraphics[height=\figheight,width=\figwidth,angle=0]{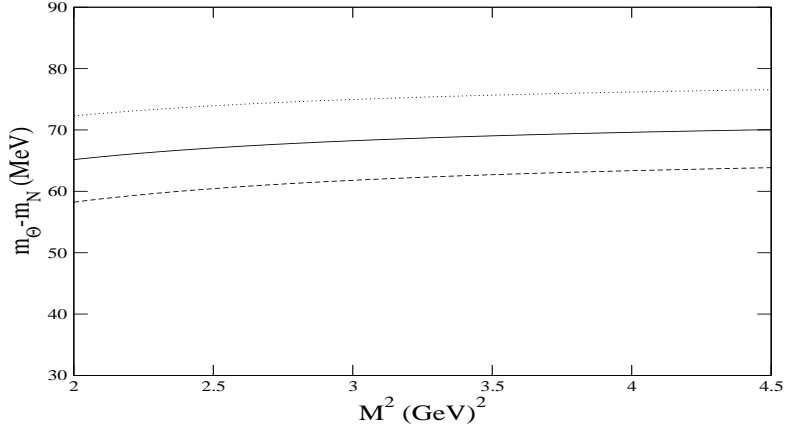}
\caption{\label{fig:3} Mass difference $m_\Theta-m_N$ for different
values of the continuum threshold, the solid, dashed and dotted lines
are for $s_{0\Theta}=3.5\ \gev^2$ and $s_{0N}=3.2\ \gev^2$,
$s_{0\Theta}=4.1\ \gev^2$ and $s_{0N}=3.8\ \gev^2$ and
$s_{0\Theta}=3.0\ \gev^2$ and $s_{0N}=2.7\ \gev^2$, respectively.}
\end{center}
\end{figure}
Similar as it has been done in \cite{Matheus:2003xr}, in fig.
\ref{fig:3} we plot the mass difference $m_\Theta-m_N$ for different values of
the continuum thresholds.
The mass splitting between the pentaquark states comes out to be about 70
MeV.
The error represented in fig. \ref{fig:3} is based on the assumption
that the continuum thresholds have the same offset for both pentaquark states.
Phenomenologically, these values can be different and one should add to the
error a part of the uncertainty from $s_0$ given in fig. \ref{fig:1}. 
Thus the error can easily amount to 50 MeV. Though the mass difference is 
consistent with the interpretation of the $N(1440)$ as a pentaquark,
the uncertainty remains large and a reduction of the error would be essential
to clarify the situation.
 
Recently, in \cite{Kondo:2004cr} it has been argued
that one should subtract all possible
colour-singlet meson-baryon contributions from the pentaquark current. We
believe that this claim is not correct. Nothing is wrong to use the current of
eq. \eqn{eq:Current}. This current contains also 2-particle 
intermediate states which
have to be added to the phenomenological side. 
However, at energies around the pentaquark mass we expect the pentaquark
contribution to dominate the spectral density. Apart from $KN$ production whose
threshold lies somewhat below the pentaquark energy other intermediate states
start at higher energy. Therefore it is expected that the baryon-meson 
continuum contribution only becomes important at energies much above the
pentaquark mass.
In this energy range the spectral density is suppressed by the exponential in
eq. \eqn{eq:TheoMoments} and the correlator should be well 
approximated by the assumption of quark-hadron duality. 
Furthermore, the current $\eta(x)$ is based on the assumption of diquark
formation. Subtracting partial contributions from the OPE side changes the
pentaquark current and can remove contributions relevant for the diquark
formation. Thus these contributions can form an important part 
of the pentaquark and should not be subtracted.


To summarise, we have performed a QCD analysis based on the approach by Jaffe
and Wilczek. We obtain a sum rule that is stable over the Borel parameter
$M$ and reproduces the mass of the pentaquark within errors. 
The error is to a large part due to the lack of experimental information above the
pentaquark energy. Furthermore, a complete calculation at next-to-leading order
would help to quantify the uncertainties in the theoretical expansion. However,
with the complex structure of the current and given the fact that this 
includes a calculation of five loops, this is a difficult task. 
We have also performed an analysis for the pentaquark with the quantum
numbers of the nucleon and have shown that the interpretation of the 
Roper resonance $N(1440)$ as $(ud)^2\bar{d}$ pentaquark state is 
consistent with the sum rules.
It is important to note that the sum rules are directly based on QCD and thus,
apart from the structure of the current, do not contain further model
assumptions. It would be interesting to see if lattice calculations could
confirm these findings. 
First lattice calculations exist \cite{Lattice} 
which, however, are based on
different interpolating currents and whose results are not yet conclusive.
Further advance in two directions seems feasible:
higher lying pentaquark states with different quantum numbers and internal
structure could be investigated and a QCD analysis based on the approach by
Karliner and Lipkin should be done. This might help to understand
the specific features of the models and to differentiate between the
approaches.


\section*{Acknowledgments}
I would like thank Antonio Pich for numerous discussions and reading the
manuscript. I thank the European Union for financial 
support under contract no. HPMF-CT-2001-01128. 
This work has been supported in part by 
EURIDICE, EC contract no. HPRN-CT-2002-00311 and by MCYT (Spain) under grant 
FPA2001-3031.


\begin{thebibliography}{99}

\bibitem{Nakano:2003qx}
T.~Nakano {\it et al.}  [LEPS Collaboration],
Phys.\ Rev.\ Lett.\  {\bf 91} (2003) 012002
[arXiv:hep-ex/0301020].

\bibitem{Barmin:2003vv}
V.~V.~Barmin {\it et al.}  [DIANA Collaboration],
Phys.\ Atom.\ Nucl.\  {\bf 66} (2003) 1715
[Yad.\ Fiz.\  {\bf 66} (2003) 1763]
[arXiv:hep-ex/0304040].

\bibitem{Stepanyan:2003qr}
S.~Stepanyan {\it et al.}  [CLAS Collaboration],
Phys.\ Rev.\ Lett.\  {\bf 91} (2003) 252001
[arXiv:hep-ex/0307018].

\bibitem{Barth:2003ja}
J.~Barth {\it et al.}  [SAPHIR Collaboration],
Phys.\ Lett.\ B {\bf 572} (2003) 127.

\bibitem{Kubarovsky:2003fi}
V.~Kubarovsky {\it et al.}  [CLAS Collaboration],
Phys.\ Rev.\ Lett.\  {\bf 92} (2004) 032001
(Erratum-ibid.\  {\bf 92} (2004) 049902)
[arXiv:hep-ex/0311046].

\bibitem{Airapetian:2003ri}
A.~Airapetian {\it et al.}  [HERMES Collaboration],
Phys.\ Lett.\ B {\bf 585} (2004) 213
[arXiv:hep-ex/0312044].

\bibitem{Asratyan:2003cb}
A.~E.~Asratyan, A.~G.~Dolgolenko and M.~A.~Kubantsev,
arXiv:hep-ex/0309042.

\bibitem{Aleev:2004sa}
A.~Aleev {\it et al.}  [SVD Collaboration],
arXiv:hep-ex/0401024.

\bibitem{:2004kn}
  [ZEUS Collaboration],
arXiv:hep-ex/0403051.

\bibitem{Abdel-Bary:2004ts}
M.~Abdel-Bary {\it et al.}  [COSY-TOF Collaboration],
arXiv:hep-ex/0403011.

\bibitem{Diquarks}
B.~Stech,
Phys.\ Rev.\ D {\bf 36} (1987) 975,
M.~Neubert and B.~Stech,
Phys.\ Lett.\ B {\bf 231} (1989) 477,
M.~Anselmino, E.~Predazzi, S.~Ekelin, S.~Fredriksson and D.~B.~Lichtenberg,
Rev.\ Mod.\ Phys.\  {\bf 65} (1993) 1199.

\bibitem{Karliner:2003sy}
M.~Karliner and H.~J.~Lipkin,
arXiv:hep-ph/0307243.

\bibitem{Jaffe:2003sg}
R.~L.~Jaffe and F.~Wilczek,
Phys.\ Rev.\ Lett.\  {\bf 91} (2003) 232003
[arXiv:hep-ph/0307341].

\bibitem{Jaffe:2004zg}
R.~Jaffe and F.~Wilczek,
arXiv:hep-ph/0401034.

\bibitem{Jennings:2003wz}
B.~K.~Jennings and K.~Maltman,
Phys.\ Rev.\ D {\bf 69} (2004) 094020
[arXiv:hep-ph/0308286].

\bibitem{SRbasis}
M.~A.~Shifman, A.~I.~Vainshtein and V.~I.~Zakharov,
Nucl.\ Phys.\ B {\bf 147} (1979) 385,
Nucl.\ Phys.\ B {\bf 147} (1979) 448,
L.~J.~Reinders, H.~Rubinstein and S.~Yazaki,
Phys.\ Rept.\  {\bf 127} (1985) 1,
S.~Narison,
World Sci.\ Lect.\ Notes Phys.\  {\bf 26} (1989) 1.

\bibitem{SRbaryons}
B.~L.~Ioffe,
Nucl.\ Phys.\ B {\bf 188} (1981) 317
[Erratum-ibid.\ B {\bf 191} (1981) 591],
Z.\ Phys.\ C {\bf 18} (1983) 67,
Y.~Chung, H.~G.~Dosch, M.~Kremer and D.~Schall,
Phys.\ Lett.\ B {\bf 102} (1981) 175,
Nucl.\ Phys.\ B {\bf 197} (1982) 55,
H.~G.~Dosch, M.~Jamin and S.~Narison,
Phys.\ Lett.\ B {\bf 220} (1989) 251.

\bibitem{SRdiquarks}
H.~G.~Dosch, M.~Jamin and B.~Stech,
Z.\ Phys.\ C {\bf 42} (1989) 167,
M.~Jamin and M.~Neubert,
Phys.\ Lett.\ B {\bf 238} (1990) 387.


\bibitem{Sugiyama:2003zk}
J.~Sugiyama, T.~Doi and M.~Oka,
Phys.\ Lett.\ B {\bf 581} (2004) 167
[arXiv:hep-ph/0309271].

\bibitem{Zhu:2003ba}
S.~L.~Zhu,
Phys.\ Rev.\ Lett.\  {\bf 91} (2003) 232002
[arXiv:hep-ph/0307345].

\bibitem{Matheus:2003xr}
R.~D.~Matheus, F.~S.~Navarra, M.~Nielsen, R.~Rodrigues da Silva and S.~H.~Lee,
Phys.\ Lett.\ B {\bf 578} (2004) 323
[arXiv:hep-ph/0309001].

\bibitem{Huang:2003bu}
P.~Z.~Huang, W.~Z.~Deng, X.~L.~Chen and S.~L.~Zhu,
Phys.\ Rev.\ D {\bf 69} (2004) 074004
[arXiv:hep-ph/0311108].

\bibitem{Pentaapproaches}
K.~Cheung,
arXiv:hep-ph/0308176,
I.~M.~Narodetskii, Y.~A.~Simonov, M.~A.~Trusov and A.~I.~Veselov,
Phys.\ Lett.\ B {\bf 578} (2004) 318
[arXiv:hep-ph/0310118],
E.~Shuryak and I.~Zahed,
arXiv:hep-ph/0310270,
F.~E.~Close,
arXiv:hep-ph/0311087,
J.~J.~Dudek,
arXiv:hep-ph/0403235,
N.~I.~Kochelev, H.~J.~Lee and V.~Vento,
arXiv:hep-ph/0404065.

\bibitem{Novikov:1985gd}
V.~A.~Novikov, M.~A.~Shifman, A.~I.~Vainshtein and V.~I.~Zakharov,
Fortsch.\ Phys.\  {\bf 32} (1985) 585.

\bibitem{Yang:1993bp}
K.~C.~Yang, W.~Y.~P.~Hwang, E.~M.~Henley and L.~S.~Kisslinger,
Phys.\ Rev.\ D {\bf 47} (1993) 3001.

\bibitem{inputparameters}
D.~J.~Broadhurst, P.~A.~Baikov, V.~A.~Ilyin, J.~Fleischer, O.~V.~Tarasov and V.~A.~Smirnov,
Phys.\ Lett.\ B {\bf 329} (1994) 103
[arXiv:hep-ph/9403274],
H.~G.~Dosch and S.~Narison,
Phys.\ Lett.\ B {\bf 417} (1998) 173
[arXiv:hep-ph/9709215],
M.~Jamin,
Phys.\ Lett.\ B {\bf 538} (2002) 71
[arXiv:hep-ph/0201174],
T.~W.~Chiu and T.~H.~Hsieh,
Nucl.\ Phys.\ B {\bf 673} (2003) 217
[Nucl.\ Phys.\ Proc.\ Suppl.\  {\bf 129} (2004) 492]
[arXiv:hep-lat/0305016].
A.~Di Giacomo and Y.~A.~Simonov,
arXiv:hep-ph/0404044.

\bibitem{Kondo:2004cr}
Y.~Kondo, O.~Morimatsu and T.~Nishikawa,
arXiv:hep-ph/0404285.

\bibitem{Lattice}
F.~Csikor, Z.~Fodor, S.~D.~Katz and T.~G.~Kovacs,
JHEP {\bf 0311} (2003) 070
[arXiv:hep-lat/0309090],
S.~Sasaki,
arXiv:hep-lat/0310014,
T.~W.~Chiu and T.~H.~Hsieh,
arXiv:hep-ph/0403020.

\end{thebibliography}

\end{document}